\documentclass{PoS}
\usepackage{amsmath,cite,epsfig,euler}
\newcommand{\Equation}[1]{Eq.~(\ref{#1})}
\newcommand{\Figure}[1]{Fig.~\ref{#1}}
\newcommand{\ie}{{\it i.e.}}
\newcommand{\cf}{{\it cf.}}

\newcommand{\srac}[2]{{\textstyle\frac{#1}{#2}}}
\newcommand{\xone}{x_1}
\newcommand{\xtwo}{x_2}
\newcommand{\lid}[2]{#1\!\cdot\!#2}
\newcommand{\bram}[1]{\langle#1|}

\newcommand{\ketm}[1]{|#1]}

\newcommand{\slashp}{p\hspace{-6.5pt}/}
\newcommand{\slashl}{\ell\hspace{-6.0pt}/}
\newcommand{\slashE}{E\hspace{-6.0pt}/}
\newcommand{\imag}{\mathrm{i}}
\newcommand{\GeV}{\mathrm{GeV}}
\newcommand{\TeV}{\mathrm{TeV}}
\newcommand{\gqcd}{g_{\mathrm{s}}}%{\mathrm{g}}

\title{Scattering amplitudes for high-energy factorization}

\ShortTitle{Scattering amplitudes for high-energy factorization}

\author{\speaker{A.~van~Hameren}\\
        The H.\ Niewodnicza\'nski Institute of Nuclear Physics\\
        Polisch Academy of Sciences\\
        Radzikowskiego 152, 31-342 Cracow, Poland\\
        E-mail: \email{hameren@ifj.edu.pl}}

\abstract{A prescription is presented to construct manifestly gauge invariant tree-level scattering amplitudes with one or two off-shell initial-state gluons for processes with arbitrary particles in the final state, which allows for calculations that are efficient and easy to automate. These scattering amplitudes are relevant for factorization schemes beyond collinear factorization that allow the gluons entering the partonic cross section to have non-vanishing transversal momentum components.}

\FullConference{XXI International Workshop on Deep-Inelastic Scattering and Related Subject -DIS2013,\\
		22-26 April 2013\\
		Marseilles,France}

\begin{document}

\section{Introduction}
The very high energy at which the Large Hadron Collider operates allows for scattering processes in which the energy entering the hard scattering process, while being still large enough for perturbative QCD to be applicable, is much smaller than the energy of the initial colliding protons and/or heavy ions.
For such scattering processes factorization procedures other than collinear factorization may be applicable, which go under the name of {\em high-energy factorization\/}.
An explicit example, which will serve as the basis of our discussion, was introduced in~\cite{Catani:1994sq}.

One of the main features of this factorization approach is that it, already at the lowest perturbative order, introduces kinematical effects in the hard matrix elements that only appear at higher order collinear factorization.
This happens through the explicit occurrence of the transversal momenta of the initial-state partons entering the parton-level scattering process.
Since the initial-state partons within high-energy factorization further carry only one longitudinal momentum component, these partons are off-shell and have negative virtuality.
This immediately raises the question how to define the hard scattering matrix element properly.
In order to be well-defined, it must be gauge-invariant and satisfy the necessary Ward identities.
Furthermore, it must preferably be calculable in different gauges, \ie\ with different choices for the gluon propagator.
It is not {\it a-priori\/} clear that this is possible if the partonic process under consideration involves off-shell partons.

One approach to solve this issue for off-shell gluons is via Lipatov's effective action~\cite{Lipatov:1995pn,Antonov:2004hh}.
In this approach, the standard QCD Lagrangian is extended with terms involving so-called reggeized gluons, and including all positive powers of the QCD coupling constant.
Gauge invariant scattering amplitudes can be expressed in terms of effective reggeon-gluon vertices.
Another approach suitable for processes with one off-shell gluon was presented in~\cite{vanHameren:2012uj}.

In this write-up, we present an approach which, instead of an effective action, requires the introduction of two auxiliary quarks and anti-quarks, accompanied with eikonal Feynman rules.
The extra terms added to the standard QCD Lagrangian are just linear in the coupling constant, and scattering amplitudes can calculated following familiar local Feynman rules.
The latter ensures that well-established efficient and automated methods for the calculation of tree-level scattering amplitudes~\cite{Mangano:2002ea,Kanaki:2000ey,Moretti:2001zz,Gleisberg:2008fv,Kleiss:2010hy} can directly be applied to calculate tree-level amplitudes with off-shell initial state gluons.
We will treat the case of two off-shell initial-state gluons, but the formalism can straightforwardly be restricted to a single off-shell initial-state gluon.

\section{Construction of gauge-invariant scattering amplitudes}
In high-energy factorization, each of the initial-state partons carries a single longitudinal momentum component, associated with one of the initial protons and/or heavy ions in the scattering process, and transverse momentum components:
%
%%%%%%%%%%%%%%%%%%%%%%%%%%%%%%%%%%%%%%%%
\begin{equation}
k_1^\mu = \xone \ell_1^\mu + k_{1\perp}^\mu
\quad,\quad
k_2^\mu = \xtwo \ell_2^\mu + k_{2\perp}^\mu
\label{eqn:k1k2}
\end{equation}
%%%%%%%%%%%%%%%%%%%%%%%%%%%%%%%%%%%%%%%%
%
where we take
%
%%%%%%%%%%%%%%%%%%%%%%%%%%%%%%%%%%%%%%%%
\begin{equation}
\ell_1=\big(\srac{E}{2},0,0,\srac{E}{2}\big)
\quad,\quad
\ell_2=\big(\srac{E}{2},0,0,-\srac{E}{2}\big)
\quad,\quad
\end{equation}
%%%%%%%%%%%%%%%%%%%%%%%%%%%%%%%%%%%%%%%%
%
and where $E$ is the total energy.
The transversal momenta satisfy $\lid{k_{1,2\perp}}{\ell_{1,2}}=0$.
The longitudinal momentum fractions $x_{1,2}$ are positive but smaller than $1$.
In order to arrive at a gauge invariant scattering amplitude for the process
%
%%%%%%%%%%%%%%%%%%%%%%%%%%%%%%%%%%%%%%%%
\begin{equation}
g^*g^*\to X
~,
\label{offshellprocess}
\end{equation}
%%%%%%%%%%%%%%%%%%%%%%%%%%%%%%%%%%%%%%%%
%
where $X$ stands for an arbitrary parton-level final state, it can be imagined to be embedded in a larger parton-level scattering process
%
%%%%%%%%%%%%%%%%%%%%%%%%%%%%%%%%%%%%%%%%
\begin{equation}
q_Aq_B\to q_Aq_B\,X
\label{embedding}
\end{equation}
%%%%%%%%%%%%%%%%%%%%%%%%%%%%%%%%%%%%%%%%
%
where $q_A,q_B$ are the auxiliary quarks mentioned in the introduction.
%
%%%%%%%%%%%%%%%%%%%%%%%%%%%%%%%%%%%%%%%%
\begin{figure}
\begin{center}
\epsfig{figure=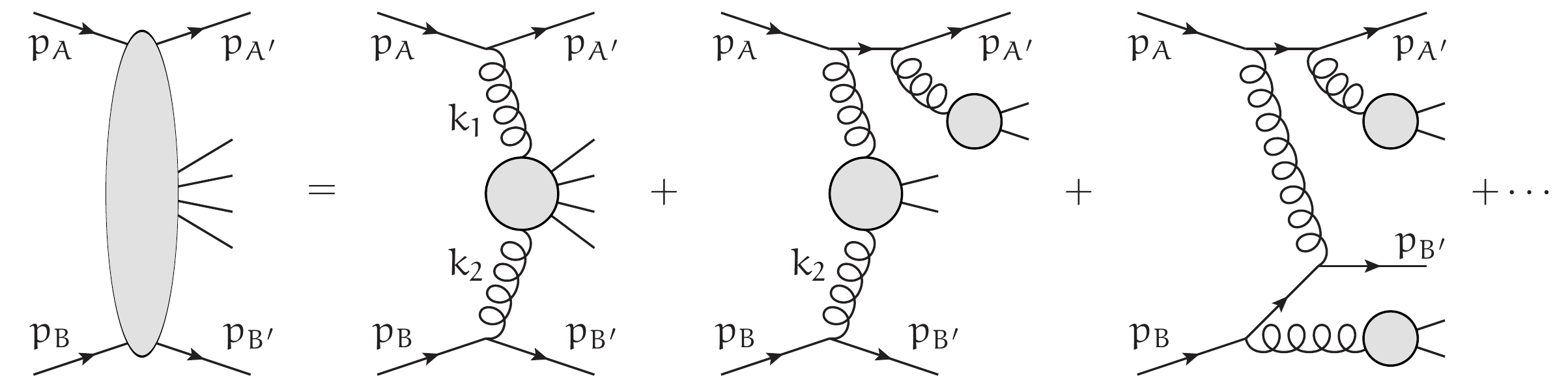,width=0.94\linewidth}
\caption{\label{Fig1}The embedding of the parton-level process $g^*g^*\to X$ into $q_A\,q_B\to q_A\,q_B\,X$.}
\end{center}
\end{figure}
%%%%%%%%%%%%%%%%%%%%%%%%%%%%%%%%%%%%%%%%
%
The Feynman graphs contributing to this process can be categorized such that one set contains all the graphs that one would naturally associate with \Equation{offshellprocess}.
This categorization is depicted in \Figure{Fig1}, and the aforementioned set is the first term on the r.h.s., and consists of all graphs containing both the propagators with $k_1$ and $k_2$.
In the collinear limit of $k_{1,2\perp}^2\to0$ this term clearly dominates the other terms, and contains exactly the Feynman graphs one would take into account in a collinear calculation.
For non-vanishing $k_{1,2\perp}^2$, however, all terms must be taken into account.

For our purpose, the momenta $p_A,p_{A'},p_B,p_{B'}$ of the auxiliary quarks need to be on-shell and need to satisfy
%
%%%%%%%%%%%%%%%%%%%%%%%%%%%%%%%%%%%%%%%%
\begin{equation}
p_A^\mu-p_{A'}^\mu=k_1^\mu\quad,\quad p_B^\mu-p_{B'}^\mu=k_2^\mu
\label{eqn:pApBk1k2}
~.
\end{equation}
%%%%%%%%%%%%%%%%%%%%%%%%%%%%%%%%%%%%%%%%
%
In order for this and \Equation{eqn:k1k2}, to hold, $p_A$ cannot be just proportional to $\ell_1$ and $p_B$ just proportional to $\ell_2$.
An elegant construction for this kinematical situation is the following.
We introduce a pair of complex four-momenta to span the transversal space
%
%%%%%%%%%%%%%%%%%%%%%%%%%%%%%%%%%%%%%%%%
\begin{equation}
\ell_3^\mu=\srac{1}{2}\bram{\ell_2}\,\gamma^\mu\,\ketm{\ell_1}
\quad,\quad
\ell_4^\mu=\srac{1}{2}\bram{\ell_1}\,\gamma^\mu\,\ketm{\ell_2}
~,
\label{eqn:l3l4}
\end{equation}
%%%%%%%%%%%%%%%%%%%%%%%%%%%%%%%%%%%%%%%%
%
so that $\lid{\ell_{1,2}}{\ell_{3,4}}=0$ and $\lid{\ell_3}{\ell_4}=-\lid{\ell_1}{\ell_2}$, and
%
%%%%%%%%%%%%%%%%%%%%%%%%%%%%%%%%%%%%%%%%
\begin{equation}
k_{1\perp}^\mu =
-\frac{\lid{k_{1\perp}}{\ell_4}}{\lid{\ell_1}{\ell_2}}\,\ell_3^\mu
-\frac{\lid{k_{1\perp}}{\ell_3}}{\lid{\ell_1}{\ell_2}}\,\ell_4^\mu
\quad,\quad
k_{2\perp}^\mu =
-\frac{\lid{k_{2\perp}}{\ell_4}}{\lid{\ell_1}{\ell_2}}\,\ell_3^\mu
-\frac{\lid{k_{2\perp}}{\ell_3}}{\lid{\ell_1}{\ell_2}}\,\ell_4^\mu
~.
\end{equation}
%%%%%%%%%%%%%%%%%%%%%%%%%%%%%%%%%%%%%%%%
%
Then, by choosing
%
%%%%%%%%%%%%%%%%%%%%%%%%%%%%%%%%%%%%%%%%
\begin{align}
p_A^\mu = (\Lambda+\xone)\ell_1^\mu
- \frac{\lid{k_{1\perp}}{\ell_4}}{\lid{\ell_1}{\ell_2}}\,\ell_3^\mu
\quad&,\quad
p_{A'}^\mu = \Lambda\ell_1^\mu
+ \frac{\lid{k_{1\perp}}{\ell_3}}{\lid{\ell_1}{\ell_2}}\,\ell_4^\mu
\notag\\
p_B^\mu = (\Lambda+\xtwo)\ell_2^\mu
- \frac{\lid{k_{2\perp}}{\ell_3}}{\lid{\ell_1}{\ell_2}}\,\ell_4^\mu
\quad&,\quad
p_{B'}^\mu = \Lambda\ell_2^\mu
+ \frac{\lid{k_{2\perp}}{\ell_3}}{\lid{\ell_1}{\ell_2}}\,\ell_3^\mu
\label{eqn:pApB}
\end{align}
%%%%%%%%%%%%%%%%%%%%%%%%%%%%%%%%%%%%%%%%
%
onshellness of $p_A,p_{A'},p_B,p_{B'}$ as well as \Equation{eqn:pApBk1k2} and \Equation{eqn:k1k2} are satisfied, for any value of the free parameter $\Lambda$.
So by evaluating the scattering amplitude for \Equation{embedding} with the momenta above, we have, despite that they are complex, a gauge invariant object~\cite{Britto:2005fq} with the desired momenta for the off-shell gluons.
Using a definition for spinors that is consistent also for complex momenta, \cf~\cite{vanHameren:2005ed}, we see that
%
%%%%%%%%%%%%%%%%%%%%%%%%%%%%%%%%%%%%%%%%
\begin{equation}
\ketm{p_A}\propto\ketm{\ell_1}
\quad,\quad \bram{p_{A'}}\propto\bram{\ell_1}
\qquad,\qquad
\ketm{p_B}\propto\ketm{\ell_2}
\quad,\quad \bram{p_{B'}}\propto\bram{\ell_2}
\label{eqn:pABspin}
% \brap{p_{B'}}\propto\brap{\ell_1} \quad&,\quad \ketp{p_B}\propto\ketp{\ell_1}
% \notag
~,
\end{equation}
%%%%%%%%%%%%%%%%%%%%%%%%%%%%%%%%%%%%%%%%
%
and that we may assign the spinors on the right-hand-sides in the relations above instead of the ones on the left-hand-sides to the external auxiliary quarks, without spoiling gauge invariance.
Still, the constructed amplitude depends on unphysical imaginary momentum components, but this dependence can now be removed by taking $\Lambda\to\infty$.
This limit only affects the auxiliary quark lines, and results in the propagators to become eikonal propagators: for a propagator on the $A$-quark line with momentum $p$ we have
%%%%%%%%%%%%%%%%%%%%%%%%%%%%%%%%%%%%%%%%
\begin{equation}
\frac{\imag\,\slashp}{p^2}
\overset{\Lambda\to\infty}{\longrightarrow}
\frac{\imag\,\slashl_1}{2\,\lid{\ell_1}{p}}
~.
\end{equation}
%%%%%%%%%%%%%%%%%%%%%%%%%%%%%%%%%%%%%%%%
%
Notice that components of $\ell_1$ and $k_{1\perp}$ in $p$ are eliminated in the propagator, and may be chosen arbitrarily.
The same works quite analogously for the $B$-quark line and $\ell_2$.

The construction above leads to the following prescription to calculate manifestly gauge invariant scattering amplitudes for \Equation{offshellprocess}:
\begin{enumerate}
\item Consider the embedding \Equation{embedding} with momentum flow as if the momenta $p_A,p_B$ of the initial-state quarks and $p_{A'},p_{B'}$ of the final-state quarks are given by
\begin{equation}
p_A^\mu=k_1^\mu
\quad,\quad
p_B^\mu=k_2^\mu
\quad,\quad
p_{A'}^\mu=p_{B'}^\mu=0
~.
\nonumber
\end{equation}
\item
Assign the spinors $\ketm{\ell_1},\bram{\ell_1}$ the external $A$-quarks, and assign $\imag\,\slashl_1/(2\lid{\ell_1}{p})$ instead of $\imag\,\slashp/p^2$ to the propagators on the $A$-quark line.
\item
Do the same with the $B$ quark line, using $\ell_2$ instead of $\ell_1$.
\item Multiply the amplitude with
$\displaystyle
\frac{\xone\sqrt{-k_{1\perp}^2}}{\sqrt{2}\,\gqcd}
\times
\frac{\xtwo\sqrt{-k_{2\perp}^2}}{\sqrt{2}\,\gqcd}
~.
$
\end{enumerate}
For the rest, normal Feynman rules apply.
This holds for color too; the auxiliary quarks are in the fundamental representation.
The factor in the last point assures the correct collinear limit when $k_{1\perp}^2,k_{2\perp}^2\to0$.

It is possible to elaborate the auxiliary quark lines, and arrive at simplified  vertices on these lines while simultaneously reducing the numerator of the eikonal propagators to $1$~\cite{vanHameren:2012if}.
The formulation presented above is however simpler, and more close to familiar Feynman rules.
It leads to the factor in point 4 which is slightly different from the one derived in~\cite{vanHameren:2012if}.
In \cite{vanHameren:2012if} it is shown that the tree-level scattering amplitudes obtained this way are equivalent to those obtained with Lipatov's effective action approach.

\section{Application}
The prescription has been implemented into a numerical program that calculates scattering amplitudes in an efficient recursive way, in the spirit of~\cite{Mangano:2002ea,Kanaki:2000ey,Moretti:2001zz,Gleisberg:2008fv,Kleiss:2010hy}.
As an example calculation to prove its computational potential, we present distributions for the processes
%
%%%%%%%%%%%%%%%%%%%%%%%%%%%%%%%%%%%%%%%%
\begin{align}
gg^*&\to b\bar{b}\,\mu^+\mu^-
~,
\label{Zprocess}\\
ug^*&\to b\bar{b}\,\mu^+\nu_\mu\,d
~.
\label{Wprocess}
\end{align}
%%%%%%%%%%%%%%%%%%%%%%%%%%%%%%%%%%%%%%%%
For the first process, only contributions involving a $Z$-boson have been taken into account, and for the second only contributions involving a single $W$-boson.
%
%%%%%%%%%%%%%%%%%%%%%%%%%%%%%%%%%%%%%%%%
\begin{figure}
\begin{center}
\epsfig{figure=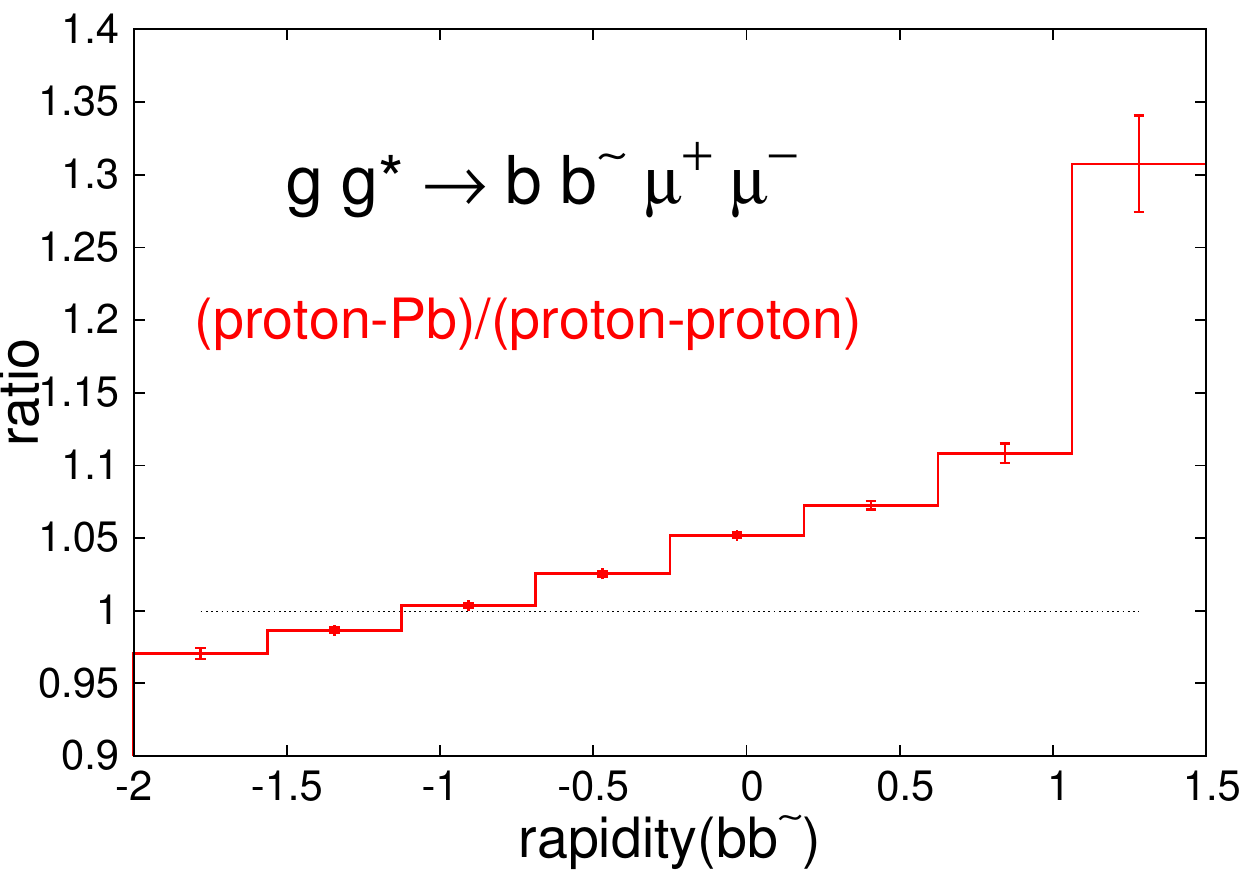,width=0.48\linewidth}
\epsfig{figure=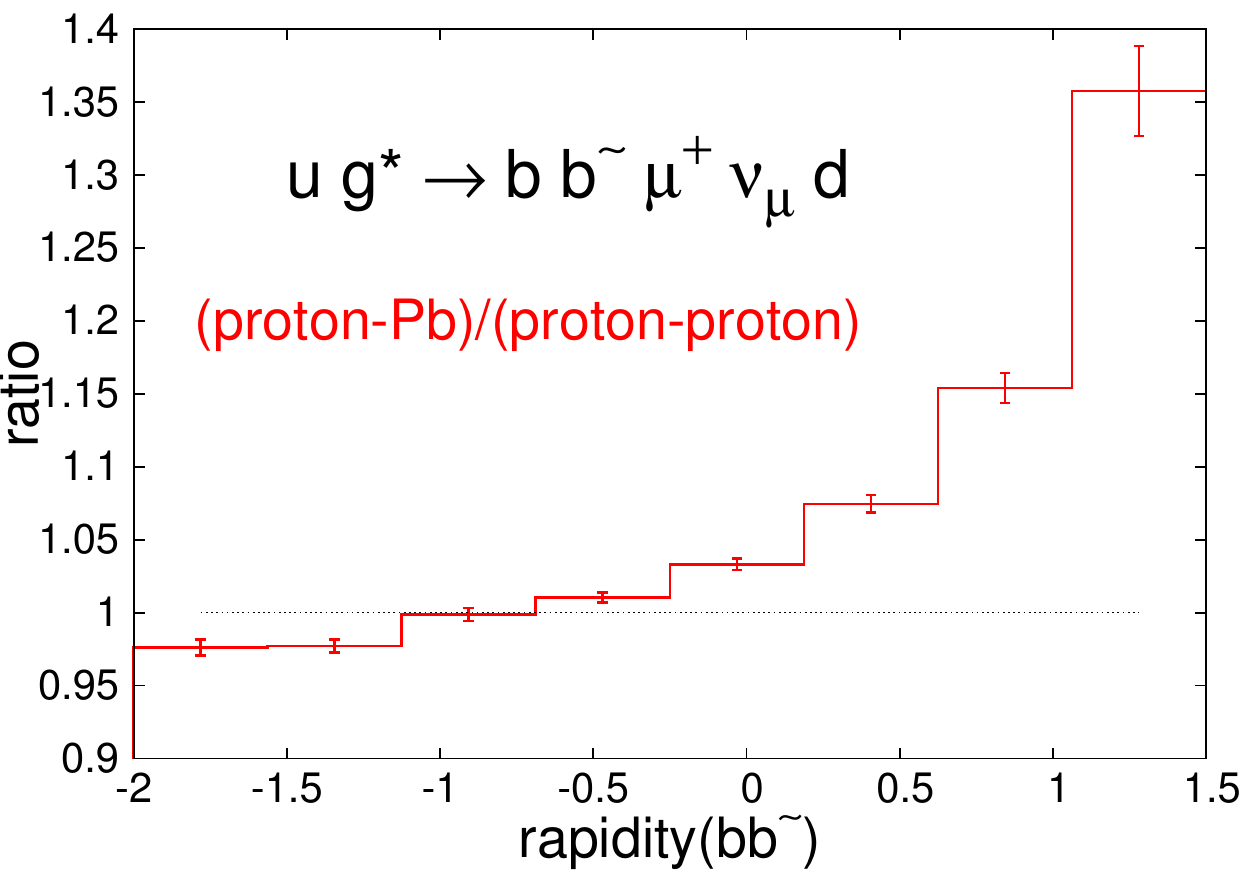,width=0.48\linewidth}
\caption{\label{Fig2}The rapidity distribution of the $b\bar{b}$ pair in the center-of-mass frame for p-Pb scattering divided by the rapidity distribution of the $b\bar{b}$ pair in the center-of-mass frame for p-p scattering.}
\end{center}
\end{figure}
%%%%%%%%%%%%%%%%%%%%%%%%%%%%%%%%%%%%%%%%
%
%
Calculations were performed at a center-off-mass energy of $5.02\TeV$.
The phase space was restricted such that $\slashE_T>20\GeV$ for the neutrino, and $p_T>20\GeV$ for all other final-state particles.
For process~(\ref{Wprocess}) we restricted the $p_T$ of the $\mu^+$ to be smaller than $50\GeV$.
Furthermore, we used rapidity cuts $|y|<2.5$ for quarks and $|y|<2.1$ for the muons, and separation cuts such that $\Delta R$ between any pair of final-state fermions, excluding the neutrino, is larger than $0.4$.

For the on-shell inital-state parton, CTEQ6.1 pdfs~\cite{Pumplin:2002vw} were used.
\Figure{Fig2} shows the ratios of the distributions of the rapidity of the $b\bar{b}$ pair for different choices of the unintegrated pdf for the off-shell gluon.
It involves the pdfs from~\cite{Kutak:2012rf} for p-Pb and p-p scattering.
The off-shell gluon has positive rapidity, and the plots reveal that the rapidity distributions for p-Pb are (in the center-of-mass frame) slightly shifted towards positive rapidities compared to the p-p distributions.
This can be understood as an indication that the p-Pb pdf is suppressed at low $x$ compared to the p-p pdf.

\section{Summary}
We presented a prescription that allows for the efficient calculation of tree-level scattering amplitudes with off-shell initial-state gluons, necessary for calculations within factorization schemes in which the initial-state gluons entering the partonic cross section carry non-vanishing transversal momentum components. It has been implemented into an explicit numerical application, and some results have been presented to proof its computational potential.

\section*{Acknowledgments}
This work was partially supported by HOMING PLUS/2010-2/6: ``Matrix Elements and Exclusive Parton Densities for Large Hadron Collider''.

%\bibliography{bibliography}{}
\providecommand{\href}[2]{#2}\begingroup\raggedright\endgroup
\bibliographystyle{JHEP}

%\begin{thebibliography}{99}
%\bibitem{...} 
%....
%
%\end{thebibliography}

\end{document}